# DOME: Recommendations for supervised machine learning validation in biology

*Ian Walsh*[1,*], *Dmytro Fishman*[2,*], *Dario Garcia-Gasulla*[3], *Tiina Titma*[4], *Gianluca Pollastri*[5], *The ELIXIR Machine Learning focus group*[#], *Jennifer Harrow*[6,+], *Fotis E. Psomopoulos*[7,+] & *Silvio C.E. Tosatto*[8,+]

[1]Bioprocessing Technology Institute, Agency for Science, Technology and Research, Singapore, [2]Institute of Computer Science, University of Tartu, Estonia, [3]Barcelona Supercomputing Center (BSC), Barcelona, Spain, [4]School of Information Technologies, Tallinn University of Technology, Estonia, [5]School of Computer Science, University College Dublin, Ireland, [6]ELIXIR HUB, South building, Wellcome Genome Campus, Hinxton, Cambridge, UK, [7]Institute of Applied Biosciences, Centre for Research and Technology Hellas, Thessaloniki, Greece, [8]Dept. of Biomedical Sciences, University of Padua, Padua, Italy.

[*]*contributed equally*
[#]*see list of co-authors at the end of the manuscript*
[+]*corresponding authors*

**Abstract**

Modern biology frequently relies on machine learning to provide predictions and improve decision processes. There have been recent calls for more scrutiny on machine learning performance and possible limitations. Here we present a set of community-wide recommendations aiming to help establish standards of supervised machine learning validation in biology. Adopting a structured methods description for machine learning based on data, optimization, model, evaluation (DOME) will aim to help both reviewers and readers to better understand and assess the performance and limitations of a method or outcome. The recommendations are formulated as questions to anyone wishing to pursue implementation of a machine learning algorithm. Answers to these questions can be easily included in the supplementary material of published papers.

**Introduction**

With the steep decline in the cost of high-throughput technologies, large amounts of biological data are being generated and made accessible to researchers. Machine learning (ML) has been brought into the spotlight as a very useful approach to understand cellular[1], genomic[2], proteomic[3], post-translational[4], metabolic[5] and drug discovery data[6] with the potential to result in ground-breaking medical applications[7,8]. This is clearly reflected in the corresponding growth of ML publications (Figure 1), reporting a wide range of modelling techniques in biology. While every novel ML method should be validated experimentally, this happens only in a fraction of the publications[9]. This sharp increase in publications inherently requires a corresponding increase in the number and depth of peer-reviews to offer critical assessment[10] and improve reproducibility[11,12].

Guidelines or recommendations on how to appropriately construct ML algorithms can help to ensure correct results and predictions[13,14]. In the biomedical research field, communities have defined standard guidelines and best practices for scientific data management[15] and reproducibility of computational tools[16,17]. On the other hand, a demand exists in the ML community for a cohesive and combined set of recommendations with respect to data, the optimization techniques, the final model, and evaluation protocols as a whole.

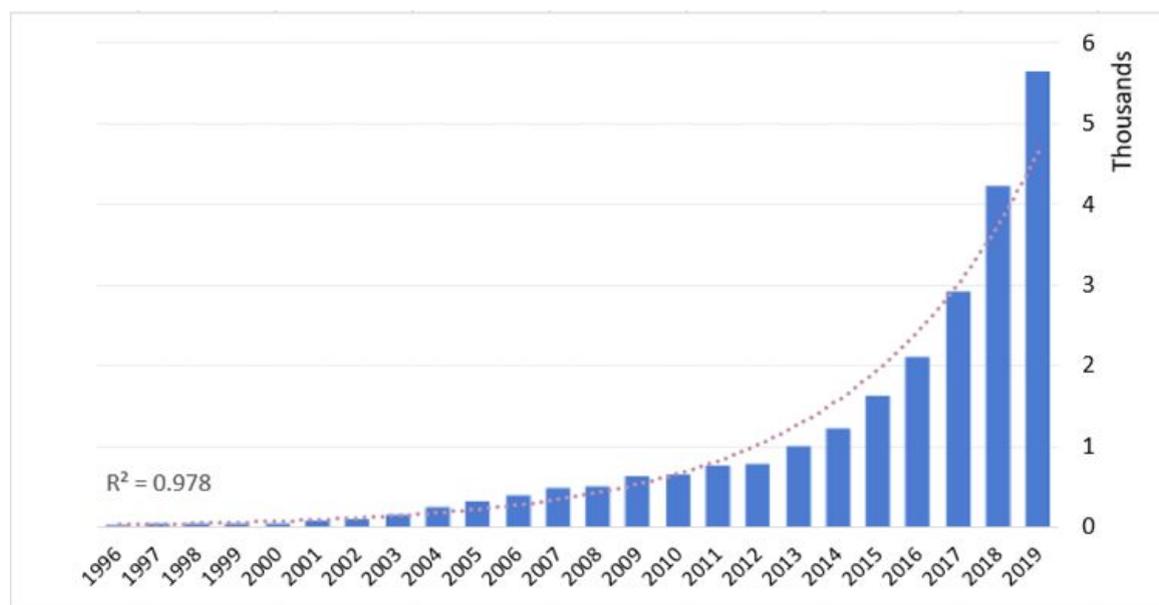

**Figure 1. Exponential increase of ML publications in biology.** The number of ML publications per year is based on Web of Science from 1996 onwards using the "topic" category for "machine learning" in combination with each of the following terms: "biolog*", "medicine", "genom*", "prote*", "cell*", "post translational", "metabolic" and "clinical".

Recently, a comment highlighted the need for standards in ML[18], arguing for the adoption of on-submission checklists[10] as a first step towards improving publication standards. Through a community-driven consensus, we propose a list of minimal requirements asked as questions to ML implementers (Box 1) that, if followed, will help to assess the quality and reliability of the proposed methods more faithfully. We have focused on Data, Optimization, Model and Evaluation (DOME) as each component of an ML implementation usually falls within one of these four topics. Importantly, no specific solutions are discussed, only recommendations (Table 1) and a checklist are provided (Box 1). Our recommendations are made primarily for the case of supervised learning in biology in the absence of direct experimental validation, as this is the most common type of ML approach used. We do not discuss how ML can be used in clinical applications[19,20]. It also remains to be investigated if the DOME recommendations can be extended to other fields of ML, like unsupervised, semi-supervised and reinforcement learning.

**Development of the recommendations**

The recommendations outlined below were initially formulated through the ELIXIR ML focus group after the publication of a comment calling for the establishment of standards for ML in biology[18]. ELIXIR, initially established in 2014, is now a mature intergovernmental European infrastructure for biological data and represents over 220 research organizations in 22 countries across many aspects of bioinformatics[21]. Over 700 national experts participate in the development and operation of national services that contribute to data access, integration, training and analysis for the research community. Over 50 of these experts involved in the field of ML have established an ELIXIR ML focus group

(https://elixir-europe.org/focus-groups/machine-learning) which held a number of meetings to develop and refine the recommendations based on a broad consensus among them.

| Broad topic | Be on the lookout for | Consequences | Recommendation(s) |
|---|---|---|---|
| **Data** | - Data size & quality<br><br>- Appropriate partitioning, dependence between train and test data.<br><br>- Class imbalance<br><br>- No access to data | ● Data not representative of domain application.<br><br>● Unreliable or biased performance evaluation.<br><br>● Cannot check data credibility. | Independence of optimization (training) and evaluation (testing) sets. (**requirement**)<br>This is especially important for meta algorithms, where independence of multiple training sets must be shown to be independent of the evaluation (testing) sets.<br><br>Release data preferably using appropriate long-term repositories, including exact splits (**requirement**)<br><br>Sufficient evidence of data size & distribution being representative of the domain. (*recommendation*) |
| **Optimization** | - Overfitting, underfitting and illegal parameter tuning<br><br>- Imprecise parameters and protocols given. | ● Reported performance too optimistic or too pessimistic.<br><br>● Models noise or miss relevant relationships.<br><br>● Results are not reproducible. | Clear statement that evaluation sets were not used for feature selection, pre-processing steps or parameter tuning. (**requirement**)<br><br>Reporting indicators on training and testing data that can aid in assessing the possibility of under/overfitting e.g. train vs. test error. (**requirement**)<br><br>Release definitions of all algorithmic hyper-parameters, regularization protocols, parameters and optimization protocol. (**requirement**)<br><br>For neural networks, release definitions of train and learning curves. (*recommendation*)<br><br>Include explicit model validation techniques, such as N-fold Cross validation. (*recommendation*) |
| **Model** | - Unclear if black box or interpretable model<br><br>- No access to: resulting source code, trained models & data<br><br>- Execution time is impractical | ● An interpretable model shows no explainable behaviour<br><br>● Cannot cross compare methods, reproducibility, & check data credibility.<br><br>● Model takes too much time to produce results | Describe the choice of black box / interpretable model. If interpretable show examples of it doing so. (**requirement**).<br><br>Release of: documented source code + models + executable + UI/webserver + software containers. (*recommendation*)<br><br>Report execution time averaged across many repeats. If computationally tough compare to similar methods (*recommendation*) |
| **Evaluation** | - Performance measures inadequate<br><br>- No comparisons to baselines or other methods | ● Biased performance measures reported.<br><br>● The method is falsely claimed as state-of-the-art. | Compare with public methods & simple models (baselines). (**requirement**)<br><br>Adoption of community validated measures and benchmark datasets for evaluation. (**requirement**)<br><br>Comparison of related methods and alternatives on the same dataset. (*recommendation*) |

| | - Highly variable performance. | • Unpredictable performance in production. | Evaluate performance on a final independent hold-out set. (*recommendation*) |
| --- | --- | --- | --- |
| | | | Confidence intervals/error intervals and statistical tests to gauge prediction robustness. (**requirement**) |

**Table 1.** Supervised ML in Biology: concerns, the consequences they impart and recommendations/requirements (recommendations in *italics* and requirements in **bold**). Key terms underlined.

---

**Box 1: Structuring a Materials and Methods Section for Supervised Machine Learning**

Here we suggest a list of questions that must be asked about each DOME section to ensure high quality of ML analysis.

- **Data**: *(this section is to be repeated separately for each dataset)*
    - *Provenance*: What is the source of the data (database, publication, direct experiment)? If data is in classes, how many data points are available in each class e.g., total for the positive ($N_{pos}$) and negative ($N_{neg}$) cases? If regression, how many real value points are there? Has the dataset been previously used by other papers and/or is it recognized by the community?
    - *Data splits*: How many data points are in the training and test sets? Was a separate validation set used, and if yes, how large was it? Is the distribution of data types in the training and test sets different? Is the distribution of data types in both training and test sets plotted?
    - *Redundancy between data splits*: How were the sets split? Are the training and test sets independent? How was this enforced (e.g. redundancy reduction to less than X% pairwise identity)? How does the distribution compare to previously published ML datasets?
    - *Availability of data*: Is the data, including the data splits used, released in a public forum? If yes, where (e.g. supporting material, URL) and how (license)?

- **Optimization**: *(this section is to be repeated separately for each trained model)*
    - *Algorithm*: What is the ML algorithm class used? Is the ML algorithm new? If yes, why is it not published in a ML journal, and why was it chosen over better known alternatives?
    - *Meta-predictions*: Does the model use data from other ML algorithms as input? If yes, which ones? Is it completely clear that training data of initial predictors and meta-predictor is independent of test data for the meta-predictor?
    - *Data encoding*: How was the data encoded and pre-processed for the ML algorithm?
    - *Parameters*: How many parameters (*p*) are used in the model? How was *p* selected?
    - *Features*: How many features (*f*) are used as input? Was feature selection performed? If yes, was it performed using the training set only?
    - *Fitting*: Is the number of parameters (*p*) much larger than the number of training points and/or is the number of features (*f*) large (e.g. in classification is $p \gg (N_{pos}+N_{neg})$ and/or $f>100$)? If yes, how was over-fitting ruled out? Conversely, if the number of training points seem very much larger than *p* and/or *f* is small (e.g. $N_{pos}+N_{neg} \gg p$ and/or $f<5$) how was under-fitting ruled out?
    - *Regularization*: were any over-fitting prevention techniques performed (e.g. early stopping using a validation set)? If yes, which ones?

- ○ *Availability of configuration*: Are the hyper-parameter configurations, optimization schedule, model files and optimization parameters reported available? If yes, where (e.g. URL) and how (license)?

- **Model**: *(this section is to be repeated separately for each trained model)*
  - ○ *Interpretability*: Is the model black box or transparent? If the model is transparent, can you give clear examples for this?
  - ○ *Output:* Is the model classification or regression?
  - ○ *Execution time*: How much real-time does a single representative prediction require on a standard machine? (e.g. seconds on a desktop PC or high-performance computing cluster)
  - ○ *Availability of software*: Is the source code released? Is a method to run the algorithm such as executable, web server, virtual machine or container instance released? If yes, where (e.g. URL) and how (license)?

- **Evaluation**:
  - ○ *Evaluation method*: How was the method evaluated? (E.g. cross-validation, independent dataset, novel experiments)
  - ○ *Performance measures*: Which performance metrics are reported? Is this set representative (e.g. compared to the literature)?
  - ○ *Comparison*: Was a comparison to publicly available methods performed on benchmark datasets? Was a comparison to simpler baselines performed?
  - ○ *Confidence*: Do the performance metrics have confidence intervals? Are the results statistically significant to claim that the method is superior to others and baselines?
  - ○ *Availability of evaluation*: Are the raw evaluation files (e.g. assignments for comparison and baselines, statistical code, confusion matrices) available? If yes, where (e.g. URL) and how (license)?

The above description is shown in table format in the Supplementary Material together with two fully worked-out examples.

---

## Scope of the recommendations

The recommendations cover four major aspects of supervised ML according to the "DOME" acronym: data, optimization, model and evaluation. The key points and rationale for each aspect of DOME is described below and summarized in Table 1. More importantly, Box 1 gives an actionable checklist to implement ML methods, with the actual recommendations codified as questions.

### 1. Data

State-of-the-art ML models are often capable of memorizing all the variation in training data. Such models when evaluated on data that they were exposed to during training would create an illusion of mastering the task at hand. However, when tested on an independent set of data (termed test or validation set), the performance would seem less impressive, suggesting low generalization power of the model. In order to tackle this problem, initial data should be divided randomly into non-overlapping parts. The simplest approach is to have independent train and test sets (possibly a third validation set). Alternatively, the cross-validation or bootstrapping techniques which choose a new train/test split multiple times from the available data, is often considered a preferred solution[22].

Overlapping of train/test data splits are particularly troublesome to overcome in biology. For example, in predictions on entire gene and protein sequences independence of train-test could be achieved by reducing the amount of homologs in the data[10,23]. Modelling enhancer-promoter contacts require a different criterion, e.g., not sharing one endpoint[24]. Modeling protein domains might require the multi-domain sequence to be split into its constituent domains before homology reduction[25]. In short, each area of biology has its own recommendations for handling overlapping data issues and previous literature is vital to put forward a strategy. In Box 1, we propose a set of questions under the section 'data splits' that should help to evaluate potential overlap between train and test data.

Reporting statistics on the data size and distribution of data types can help show if there is a good domain representation in all sets. Simple plots and/or tables showing the number of classes (classification), histogram of real values binned (regression), the different types of biological molecules in the data are vital pieces of information for each set. Further, in classification, including methods that address imbalanced classes[26,27] are also needed if the class frequencies show so. Models trained on one dataset may not be successful in dealing with data coming from adjacent but not identical datasets, a phenomenon known as covariance shift. The scale of this effect has been demonstrated in several recent publications, e.g. for prediction of disease risk from exome sequencing[28]. Although up to now the covariance shift remains an open problem, several potential solutions have been proposed in the area of transfer learning[29]. Moreover, the problem of training ML models that can generalize well on small training data, usually requires special models and algorithms[30].

Lastly, it is important to make as much data available for the public as possible[12]. Having open access to the data used for experiments including precise data splits would ensure better reproducibility of published research and as a result will improve the overall quality of published ML papers. If datasets are not readily available for example in public repositories, authors should be encouraged to find the most appropriate one, e.g. ELIXIR Deposition databases or Zenodo, to guarantee the long-term availability of such data.

## 2. Optimization

Optimization, also known as training, refers to the process of changing values that constitute the model (parameters and hyper-parameters), including pre-processing steps, in a way that maximizes the model's ability to solve a given problem. A poor choice of optimization strategy may lead to issues such as over- or under-fitting[31]. A model that has suffered severe over-fitting will show an excellent performance on training data, while performing poorly on unseen data, rendering it useless for real-life applications. On the other side of the spectrum, under-fitting occurs when very simple models capable of capturing only straightforward dependencies between features are applied to data of a more complex nature. Algorithms for feature selection[32] can be employed to reduce the chances of over-fitting. However, feature selection and other pre-processing actions come with their own recommendations. The main one being to abstain from using non-training data for feature selection and pre-processing - a particularly hard issue to spot for meta-predictors, which may lead to an overestimation of performance.

Finally, the release of files showing the exact specification of the optimization protocol and the type of parameters/hyper-parameters are a vital characteristic of the final algorithm. Lack of documentation, including limited accessibility to relevant records for the involved parameters, hyper-parameters and optimization protocol may further compound the understanding of the overall model performance.

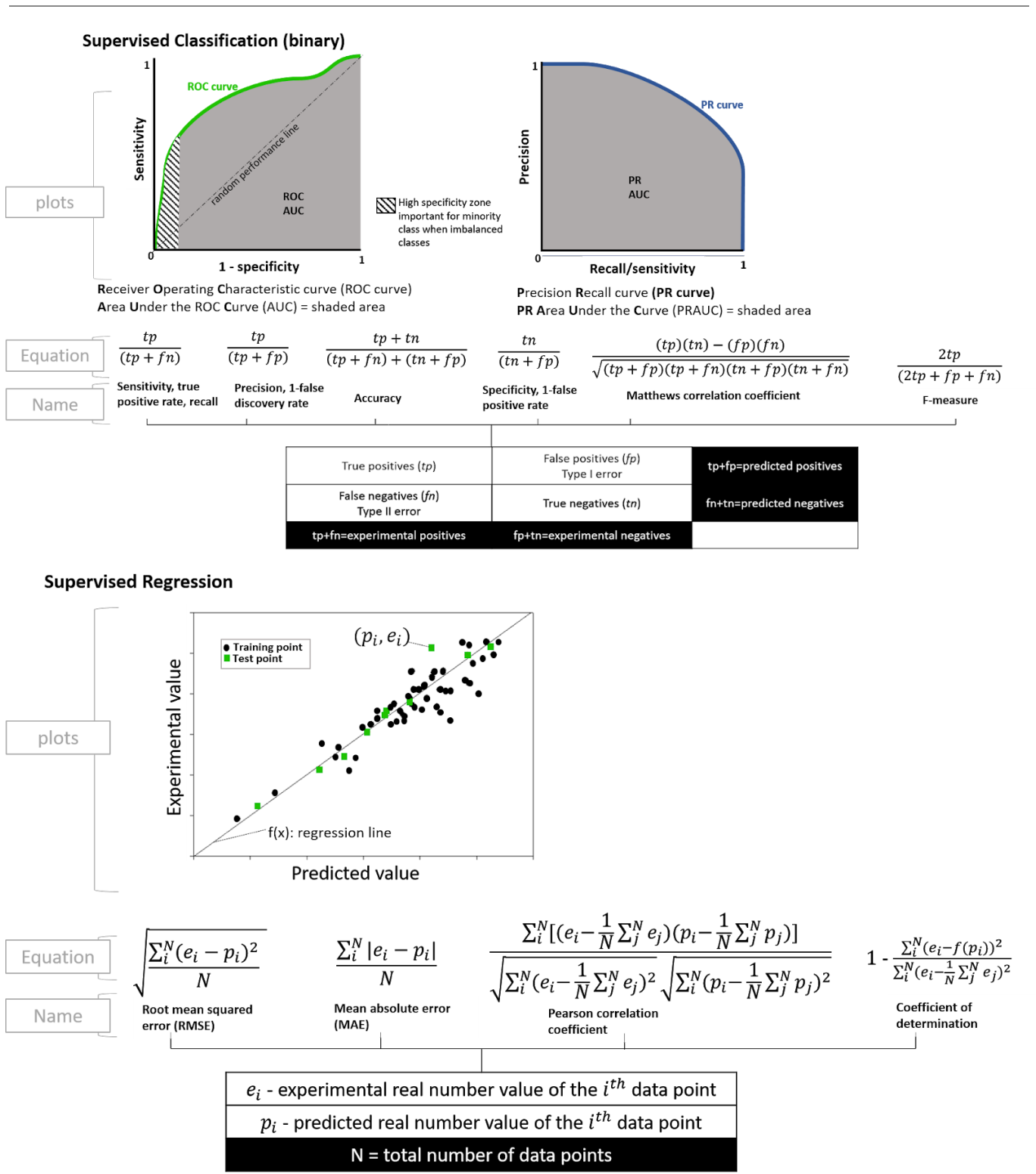

**Figure 2. (Top) Classification metrics.** For binary classification, true positives (tp), false positives (fp), false negatives (fn) and true negatives (tn) together form the confusion matrix. As all classification measures can be calculated from combinations of these four basic values, the confusion matrix should be provided as a core metric. Several measures (shown as equations) and plots should be used to evaluate the ML methods. For descriptions on how to adapt these metrics to multi-class problems see [35]. **(Bottom) Regression metrics.** ML regression attempts to produce predicted values (p) matching experimental values (e). Metrics (shown as equations) attempt to capture the difference in various ways. Alternatively, a plot can provide a visual way to represent the differences. It is advisable to report all in any ML work.

## 3. Model

Equally important aspects related to ML models are their interpretability and reproducibility. Interpretable models can infer causal relationships from the data and can output logical reasoning for each of its predictions. They are especially relevant in areas of discovery such as drug design[6] and diagnostics[33]. Conversely, black box models often give accurate predictions but do not provide insight into why they made the predictions in a way humans can understand. Both interpretable and black box models are discussed in more detail elsewhere[34]. However, developing recommendations on the choice of black box or interpretability cannot be made as both have their merits given certain situations. The main recommendation would be that there is a statement on the model type, i.e. is it black box or interpretable (see Box 1), and if it is interpretable it should be a requirement to give clear examples of it doing so.

Reproducibility is a key component to ensuring research outcomes can be further utilized and validated by the wider community. Poor model reproducibility extends beyond the documentation and reporting of the involved parameters, hyper-parameters and optimization protocol. Lacking access to the various components of a model (source code, model files, parameter configurations, executables), as well as having steep computational requirements to execute the trained models in the context of generating predictions based on new data, can both make reproducibility of the model either limited or impossible.

## 4. Evaluation

There are two types of evaluation scenarios in biological research. The first is the experimental validation of the predictions made by the ML model in the laboratory. This is highly desirable but beyond the possibilities of many ML studies. The second is a computational assessment of the model performance using established metrics. The following deals with the latter and there are a few possible related risks with computational evaluation.

Starting with the performance metrics, i.e. the quantifiable indicators of a model's ability to solve the given task, there are dozens of metrics available[35] for assessing different ML classification and regression problems. However, the plethora of options available, combined with the domain-specific expertise that might be required to select the appropriate metrics, can lead to the selection of inadequate performance measures. Often, there are critical assessment communities advocating certain performance metrics for biological ML models, e.g. CAFA[3] and CAGI[28], and it is recommended a new algorithm should use ones from the literature and critical assessments. In the absence of literature, the ones shown in Figure 2 could be a starting point.

Once performance metrics are decided, methods published in the same biological domain must be cross-compared using appropriate statistical tests (e.g. Student's t-test) and confidence intervals. Additionally, to prevent the release of ML methods that appear sophisticated but perform no better than simpler algorithms, baselines should be compared to the 'sophisticated' method and proven to be statistically inferior (e.g. small vs. deep neural networks).

## Open areas and limitations of the proposed recommendations

The field of Machine Learning in biology is vast. As such, it would be a herculean task to identify cross-cutting recommendations that could be applicable to all branches of ML across all possible domains. Therefore, this work focuses exclusively on supervised ML applications, excluding unsupervised ML, clinical applications, and individual implementations through pipelines to name a few. Moreover, the primary goal of this work is to define requirements and best practices that can be of use in the reviewing process and creation of ML-related papers, while remaining agnostic as to the actual underlying solutions. The intent is to trigger a discussion in the wider ML community leading to future work addressing possible solutions.

Several key issues related to reproducibility (e.g. data is not published, data splits are not reported and model source code with its final parameters and hyperparameters are not released) can be aided by a multitude of workflow systems that help to ensure and automate multi-step processes are completely reproducible by tracking model parameters and exact versions of the source code and libraries. Examples of commonly used workflows include Galaxy[36] and Nextflow[37]. Another *de facto* standard practice in software engineering is using version control systems such as Github to create an online copy of the source code, which can also include parameters and documentation. Similar version control systems exist for datasets. Public repositories can store experimental data on demand for significant amounts of time, enabling a long-term reproducibility of the experiment. Existing software engineering tools can be easily used to address many of the DOME recommendations.

Although having additional, more topic-specific recommendations in the future will undoubtedly be useful, in this work we aim to provide a first version that could be of general interest. Adapting the DOME recommendations to address the unique aspects of specific topics and domains, would be a task of those particular (sub-)communities. For example, having guidelines for data independence is tricky because each biological domain has its own set of guidelines for this. Nonetheless, we believe it is relevant to at least have a recommendation that authors describe how they achieved data split independence. Discussions on the correct independence strategies are needed for all of biology. Given constructive consultation processes with ML communities, relying on our own experience, it is our belief that our manuscript can be useful as a first iteration of the recommendations for supervised ML in biology. This will have the additional benefit of kickstarting the community discussion with a coherent but rough set of goals, thus facilitating the overall engagement and involvement of key stakeholders. For instance, topics to be addressed by (sub-)communities is how to adapt DOME to entire pipelines, unsupervised, semi-supervised, reinforcement, and other types of ML. E.g. in unsupervised learning, the evaluation metrics shown in Figure 2 would not apply and a completely new set of definitions are needed. Another considerable debate, as AI becomes more commonplace in society, is that ML algorithms differ in their ability to explain learned patterns back to humans. Humans naturally prefer actions or predictions to be made with reasons given. This is the black box vs. interpretability debate and we point those interested to excellent reviews as a starting point for thoughtful discussions[38–41].

Finally, we address the governance structure by suggesting a community-managed governance model similar to the open-source initiatives[42]. Community managed governance has been used in initiatives such as MIAME[43] or the PSI-MI format[44]. This sort of structure ensures continuous community consultation and improvement of the recommendations in collaboration with academic (CLAIRE, see: https://claire-ai.org/) and industrial networks (Pistoia Alliance, see: https://www.pistoiaalliance.org/). More importantly, this can be applied in particular to ML (sub-)communities working with specific problems requiring more detailed guidelines, e.g. imaging or clinical applications. We have set up a website (URL: https://www.dome-ml.org/) to provide a platform for governance and community involvement around the DOME recommendations, where news and upcoming events will be posted. As the recommendations and minimal requirements evolve over time, a version history will be available on the website. The template methods section in human (e.g. DOC) and machine readable format (YAML), as well as software for the automatic conversion of a YAML file into a human readable one are available from a dedicated GitHub repository (URL: https://github.com/MachineLearning-ELIXIR/dome-ml).

**Conclusion**

The objective of our recommendations is to increase the transparency and reproducibility of ML methods for the reader, the reviewer, the experimentalist, and the wider community. We recognize that these recommendations are not exhaustive and should be viewed as a consensus-based first iteration of a continuously evolving system of community self-review. One of the most pressing issues is to agree to a standardized data structure to describe the most relevant features of the ML methods being presented. As a first step to address this issue, we recommend including an "ML summary table", derived from Box 1, in future ML studies (see Supplementary Material). We recommend including the following sentence in the methods section of all papers: "To increase the reproducibility of the machine learning method of this study, the machine learning summary table (Table X) is included in the supporting information as per consensus guidelines (with reference to this manuscript)."

The development of a standardized approach for reporting ML methods has major advantages in increasing the quality of publishing ML methods. First, the disparity in manuscripts of reporting key elements of the ML method can make reviewing and assessing the ML method challenging. Second, certain key statistics and metrics that may affect the validity of the publication's conclusions are sometimes not mentioned at all. Third, there are unexplored opportunities associated with meta-analyses of ML datasets. Access to large sets of data can enhance both the comparison between methods and facilitate the development of better-performing methods while reducing unnecessary repetition of data generation. We believe that our recommendations to include a "machine learning summary table" and the availability of data as described above will greatly benefit the ML community and improve its standing with the intended users of these methods.

## The ELIXIR Machine Learning focus group

Emidio Capriotti (ORCID: 0000-0002-2323-0963)
Department of Pharmacy and Biotechnology, University of Bologna, Bologna (Italy)

Rita Casadio (ORCID: 0002-7462-7039)
Biocomputing Group, University of Bologna, Italy; IBIOM-CNR,Italy

Salvador Capella-Gutierrez (ORCID: 0000-0002-0309-604X)
INB Coordination Unit, Life Science Department. Barcelona Supercomputing Center (BSC), Barcelona, Spain

Davide Cirillo (ORCID: 0000-0003-4982-4716)
Life Science Department. Barcelona Supercomputing Center (BSC), Barcelona, Spain

Alexandros C. Dimopoulos (ORCID: 0000-0002-4602-2040)
Institute for Fundamental Biomedical Science, Biomedical Sciences Research Center "Alexander Fleming", Athens, Greece

Victoria Dominguez Del Angel (ORCID: 0000-0002-5514-6651)
Centre National de Recherche Scientifique, University Paris-Saclay, IFB, France

Joaquin Dopazo (ORCID: 0000-0003-3318-120X)
Clinical Bioinformatics Area, Fundación Progreso y Salud, Sevilla, Spain

Piero Fariselli (ORCID: 0000-0003-1811-4762)

Department of Medical Sciences, University of Turin, Turin, Italy

José Mª Fernández (ORCID: 0000-0002-4806-5140)
INB Coordination Unit, Life Sciences Department, Barcelona Supercomputing Center (BSC), Barcelona, Spain

Dmytro Fishman (ORCID: 0000-0002-4644-8893)
Institute of Computer Science, University of Tartu, Estonia

Dario Garcia-Gasulla (ORCID: 0000-0001-6732-5641)
Barcelona Supercomputing Center (BSC), Barcelona, Spain

Jen Harrow (ORCID:0000-0003-0338-3070)
ELIXIR HUB, South building, Wellcome Genome Campus, Hinxton, Cambridge, UK.

Florian Huber (ORCID: 0000-0002-3535-9406)
Netherlands eScience Center, Amsterdam, the Netherlands.

Anna Kreshuk (ORCID:0000-0003-1334-6388)
EMBL Heidelberg

Tom Lenaerts (ORCID: 0000-0003-3645-1455)
Machine Learning Group, Université Libre de Bruxelles, Artificial Intelligence Lab, Vrije Universiteit Brussel and Interuniversity Institute of Bioinformatics in Brussels,Brussels, Belgium.

Pier Luigi Martelli (ORCID: 0000-0002-0274-5669)
Biocomputing Group, University of Bologna, Italy

Arcadi Navarro (ORCID: 0000-0003-2162-8246)
Institute of Evolutionary Biology (Department of Experimental and Health Sciences, CSIC-Universitat Pompeu Fabra), Barcelona, Spain
Catalan Institution of Research and Advanced Studies (ICREA), Barcelona, Spain
CRG, Centre for Genomic Regulation, Barcelona Institute of Science and Technology (BIST), Barcelona, Spain

Marco Necci (ORCID: 0000-0001-9377-482X)
Dept. of Biomedical Sciences, University of Padua, Padua, Italy.

Pilib Ó Broin (ORCID: 0000-0002-6702-8564)
School of Mathematics, Statistics & Applied Mathematics, National University of Ireland Galway, Ireland

Janet Piñero (ORCID: 0000-0003-1244-7654)
Research Programme on Biomedical Informatics (GRIB), Hospital del Mar Medical Research Institute (IMIM), Department of Experimental and Health Sciences, Pompeu Fabra University (UPF), Barcelona, Spain

Damiano Piovesan (ORCID: 0000-0001-8210-2390)
Dept. of Biomedical Sciences, University of Padua, Padua, Italy.

Gianluca Pollastri (ORCID: 0000-0002-5825-4949)
School of Computer Science, University College Dublin, Ireland

Fotis E. Psomopoulos (ORCID: 0000-0002-0222-4273)
Institute of Applied Biosciences, Centre for Research and Technology Hellas, Thessaloniki, Greece

Martin Reczko (ORCID: 0000-0002-0005-8718)
Institute for Fundamental Biomedical Science, Biomedical Sciences Research Center "Alexander Fleming", Athens, Greece

Francesco Ronzano (ORCID: 0000-0001-5037-9061)
Research Programme on Biomedical Informatics (GRIB), Hospital del Mar Medical Research Institute (IMIM), Department of Experimental and Health Sciences, Pompeu Fabra University (UPF), Barcelona, Spain

Venkata Satagopam (ORCID: 0000-0002-6532-5880)
Luxembourg Centre For Systems Biomedicine (LCSB), University of Luxembourg and ELIXIR-Luxembourg Node

Castrense Savojardo (ORCID: 0000-0002-7359-0633)
Biocomputing Group, University of Bologna, Italy

Vojtech Spiwok (ORCID: 0000-0001-8108-2033)
Department of Biochemistry and Microbiology, University of Chemistry and Technology, Prague, ELIXIR-Czech Republic

Marco Antonio Tangaro (ORCID: 0000-0003-3923-2266)
Institute of Biomembranes, Bioenergetics and Molecular Biotechnologies, National Research Council (CNR), Bari, Italy

Giacomo Tartari
Institute of Biomembranes, Bioenergetics and Molecular Biotechnologies, National Research Council (CNR), Bari, Italy

David Salgado (ORCID: 0000-0002-5905-3591)
Aix Marseille Univ, INSERM, MMG UMR1251, 13005 Marseille, France.

Tiina Titma (ORCID: 0000-0002-4935-8914)
School of Information Technologies, Tallinn University of Technology, Estonia

Silvio C. E. Tosatto (ORCID: 0000-0003-4525-7793)
Dept. of Biomedical Sciences, University of Padua, Padua, Italy.

Alfonso Valencia (ORCID:0000-0002-8937-6789)
Catalan Institution of Research and Advanced Studies (ICREA), Barcelona, Spain
Life Science Department. Barcelona Supercomputing Center (BSC), Barcelona, Spain

Ian Walsh (ORCID: 0000-0003-3994-5522)
Bioprocessing Technology Institute, Agency for Science, Technology and Research, Singapore

Federico Zambelli (ORCID: 0000-0003-3487-4331)
Dept. of Biosciences, University of Milan, Milan, Italy

**Author contributions**

IW, DF, JH, FP and SCET guided the development, writing and final edits. All members of the ELIXIR machine learning focus group contributed to the discussions leading to the recommendations and writing of the manuscript.

**Competing interests**

The authors declare no competing interests.

**Acknowledgements**

The work of the ML focus group was funded by ELIXIR, the Research infrastructure for life-science data. IW was funded by Core Budget of Singapore Agency for Science Technology and Research (A*STAR).

# Supplementary Material

## Machine learning summary table:

| DOME | Version | *1.0* |
|---|---|---|
| **Data** | Provenance | *Source of data, data points (positive, $N_{pos}$ / negative, $N_{neg}$). Used by previous papers and/or community.* |
| | Dataset splits | *Size of $N_{pos}$ and $N_{neg}$ of training set, validation set (if present), test set. Distribution of $N_{pos}$ and $N_{neg}$ across sets.* |
| *(section to be repeated for each dataset)* | Redundancy between data splits | *Independence between sets. Strategy used to make examples representative (e.g. eliminating data points more similar than X%). Comparison relative to other datasets.* |
| | Availability of data | *Yes/no for datasets. If yes: Supporting information, website URL, license.* |
| **Optimization** | Algorithm | *ML class (e.g. neural network, random forest, SVM). If novel approach, reason is it not previously published.* |
| *(section to be repeated for each trained model)* | Meta-predictions | *Yes/No. If yes: how other methods are used and whether the datasets are clearly independent.* |
| | Data encoding | *How input data is transformed (e.g. global features, sliding window on sequence).* |
| | Parameters | *Number of ML model parameters (p), e.g. tunable weights in neural networks. Protocol used to select p.* |
| | Features | *Number of ML input features (f), i.e. encoding of data points. In case of feature selection: Protocol used, indicating whether it was performed on training data only.* |
| | Fitting | *Justification for excluding over- (if $p \gg N_{pos,\ train}+N_{neg,\ train}$ or f > 100) and under-fitting (if $p \ll N_{pos,\ train}+N_{neg,\ train}$ or f < 5).* |
| | Regularization | *Yes/no for overfitting prevention. If yes: specify details and parameters of technique used.* |
| | Availability of configuration | *Yes/no for hyper-parameter configuration and model files. If yes: Supporting information, website URL, license.* |
| **Model** | Interpretability | *Black box or transparent. If transparent, provide examples.* |
| *(section to be repeated for each trained model)* | Output | *Specify whether the model produces a classification (e.g. binary predictions) or regression (e.g. probability score).* |
| | Execution time | *CPU time of single representative execution on standard hardware (e.g. seconds on desktop PC).* |
| | Availability of software | *Source code repository (e.g. GitHub), software container, website URL, license.* |

| Evaluation | Evaluation method | *Cross-validation, independent dataset or novel experiments.* |
|---|---|---|
| | Performance measures | *Accuracy, sensitivity, specificity, etc.* |
| | Comparison | *Name of other methods and, if available, definition of baselines compared to. Justification of representativeness.* |
| | Confidence | *Confidence intervals and statistical significance. Justification for claiming performance differences.* |
| | Availability of evaluation | *Yes/no for raw evaluation files (e.g. assignments for comparison and baselines, confusion matrices). If yes: Supporting information, website URL, license.* |

## Example tables for author reference:

The following is an example for a primary ML summary table built from (Walsh et al., Bioinformatics 2012).

| **DOME** | Version | *1.0* |
|---|---|---|
| **Data: *X-ray disorder*** | Provenance | *Protein Data Bank (PDB) X-ray structures until May 2008 (training) and from May 2008 until September 2010 (test). 3,813 proteins total. $N_{pos}$ = 44,433 residues. $N_{neg}$ = 710,207 residues. Not previously used.* |
| | Dataset splits | *$N_{pos,train}$ = 37,495. $N_{neg,train}$ = 622,625. $N_{pos,test}$ = 6,938. $N_{neg,test}$ = 87,582 residues. No separate validation set. 5.68% positives on training set. 7.34% positives on test set.* |
| | Redundancy between data splits | *Maximum pairwise identity within and between training and test set is 25% enforced with UniqueProt tool.* |
| | Availability of data | *Yes, URL: http://protein.bio.unipd.it/espritz/.*<br>*Free use license.* |
| **Data: *DisProt disorder*** | Provenance | *DisProt version 3.7 (January 2008) for training set, DisProt version 5.7 for test set. 536 proteins total. $N_{pos}$ = 63,841 residues. $N_{neg}$ = 164,682 residues. Not previously used.* |
| | Dataset splits | *$N_{pos,train}$ = 56,414. $N_{neg,train}$ = 163,010. $N_{pos,test}$ = 7,427. $N_{neg,test}$ = 1,672 residues. No separate validation set. 25.71% positives on training set. 41.04% positives on test set.* |
| | Redundancy between data splits | *Maximum pairwise identity within and between training and test set is 40% enforced with UniqueProt tool. Less stringent threshold used to maintain adequate dataset size.* |
| | Availability of data | *Yes, URL: http://protein.bio.unipd.it/espritz/.*<br>*Free use license.* |

| **Data: *NMR disorder*** | Provenance | *Protein Data Bank (PDB) NMR structures until May 2008 (training) and from May 2008 until September 2010 (test) analyzed using the Mobi software. 2,858 proteins total. $N_{pos}$ = 40,368 residues. $N_{neg}$ = 192,170 residues. Not previously used.* |
|---|---|---|
| | Dataset splits | *$N_{pos,train}$ = 29,263. $N_{neg,train}$ = 143,891. $N_{pos,test}$ = 11,105. $N_{neg,test}$ = 48,279 residues. No separate validation set. 16.9% positives on training set. 18.7% positives on test set.* |
| | Redundancy between data splits | *Maximum pairwise identity within and between training and test set is 25% enforced with UniqueProt tool.* |
| | Availability of data | *Yes, URL: http://protein.bio.unipd.it/espritz/. Free use license.* |
| **Optimization** | Algorithm | *BRNN (Bi-directional recurrent neural network) with ensemble averaging.* |
| | Meta-predictions | *No.* |
| | Data encoding | *Sliding window of length 23 residues on input sequence with "one hot" encoding (i.e. 20 inputs per residue).* |
| | Parameters | *ESpritz p = 4,647 to 5,886 depending on model used. No optimization.* |
| | Features | *ESpritz f = 460 for sliding window of 23 residues with 20 inputs per residue. No feature selection.* |
| | Fitting | *The number of training examples is between 30 and 100 times p, suggesting neither over- nor under-fitting.* |
| | Regularization | *No. The training data is 30-100 times the number of model parameters.* |
| | Availability of configuration | *No.* |
| **Model** | Interpretability | *Black box, as correlation between input and output is masked. No attempt was made to make the model transparent.* |
| | Output | *Regression, i.e. probability of residues being disordered.* |
| | Execution time | *ESpritzS ca. 1 sec / protein, ESpritzP ca. 1,500 sec / protein on a single Intel Xeon core.* |
| | Availability of software | *Web server, URL: http://protein.bio.unipd.it/espritz/ Linux executable, URL: http://protein.bio.unipd.it/download/. Bespoke license free for academic use.* |
| **Evaluation** | Evaluation method | *Independent datasets.* |
| | Performance measures | *Accuracy, sensitivity, specificity, selectivity, F-measure, MCC, AUC are standard. $S_w$ = Sens + Spec -1.* |
| | Comparison | *Disopred, MULTICOM, DisEMBL, IUpred, PONDR-FIT, Spritz, CSpritz. Wide range of popular predictors used for comparison.* |
| | Confidence | *Bootstrapping was used to estimate statistical significance as in CASP-8 (Noivirt-Brik et al, Proteins 2009). 80% of target proteins were randomly selected 1000 times, and the standard error of the scores was* |
| | | *calculated (i.e. 1.96 × standard_error gives 95% confidence around mean for normal distributions).* |
| | Availability of evaluation | *No.* |

The following is an example for meta-predictor ML summary built from (Necci et al., Bioinformatics 2017).

| **DOME** | Version | *1.0* |
|---|---|---|
| **Data** | Provenance | *Protein Data Bank (PDB). X-ray structures missing residues. $N_{pos}$ = 339,603 residues. $N_{neg}$ = 6,168,717 residues. Previously used in (Walsh et al., Bioinformatics 2015) as an independent benchmark set.* |
| | Dataset splits | *training set: N/A*<br>*$N_{pos,test}$ = 339,603 residues. $N_{neg,test}$ = 6,168,717 residues. No validation set. 5.22% positives on the test set.* |
| | Redundancy between data splits | *Not applicable.* |
| | Availability of data | *Yes, URL: http://protein.bio.unipd.it/mobidblite/.*<br>*Free use license.* |
| **Optimization** | Algorithm | *Majority-based consensus classification based on 8 primary ML methods and post-processing.* |
| | Meta-predictions | *Yes, predictor output is a binary prediction computed from the consensus of other methods; Independence of training sets of other methods with test set of meta-predictor was not tested since datasets from other methods were not available.* |
| | Data encoding | *Label-wise average of 8 binary predictions.* |
| | Parameters | *p = 3 (Consensus score threshold, expansion-erosion window, length threshold). No optimization.* |
| | Features | *Not applicable.* |
| | Fitting | *Single input ML methods are used with default parameters. Optimization is a simple majority.* |
| | Regularization | *No.* |
| | Availability of configuration | *Not applicable.* |
| **Model** | Interpretability | *Transparent, in so far as meta-prediction is concerned. Consensus and post processing over other methods predictions (which are mostly balck boxes). No attempt was made to make the meta-prediction a black box.* |
| | Output | *Classification, i.e. residues thought to be disordered.* |

|  | Execution time | *ca. 1 second per representative on a desktop PC.* |
|---|---|---|
|  | Availability of software | *Yes, URL: http://protein.bio.unipd.it/mobidblite/.*<br>*Bespoke license free for academic use.* |
| **Evaluation** | Evaluation method | *Independent dataset* |
|  | Performance measures | *Balanced Accuracy, Precision, Sensitivity, Specificity, F1, MCC.* |
|  | Comparison | *DisEmbl-465, DisEmbl-HL, ESpritz Disprot, ESpritz NMR, ESpritz Xray, Globplot, IUPred long, IUPred short, VSL2b. Chosen methods are the methods from which the meta prediction is obtained.* |
|  | Confidence | *Not calculated.* |
|  | Availability of evaluation | *No.* |